\documentclass[twocolumn,english,aps,prb,showpacs,superscriptaddress]{revtex4-1}
\usepackage[T1]{fontenc}
\usepackage[latin9]{inputenc}
\usepackage{textcomp}
\usepackage{amsmath}
\usepackage{amssymb}
\usepackage{graphicx}
\usepackage{color}

\DeclareMathOperator{\sgn}{sgn}
\graphicspath{{figs/}{figsgaoerb/}}

\makeatletter

\usepackage{ulem}
\usepackage{float}
\usepackage{lipsum}
\usepackage{babel}

\makeatother

\usepackage{babel}
\begin{document}
\title{Engineering mobility in quasiperiodic lattices with exact mobility edges}
%\title{Engineering localization-delocalization transition in a quasiperiodic lattice with exact mobility edge}
\author{Zhenbo Wang}
\affiliation{Institute of Theoretical Physics, State Key Laboratory of Quantum Optics and Quantum Optics Devices, Collaborative Innovation Center of Extreme Optics, Shanxi University, Taiyuan 030006, P. R. China}

\author{Yu Zhang}
\affiliation{Beijing National Laboratory for Condensed Matter Physics, Institute
of Physics, Chinese Academy of Sciences, Beijing 100190, China}
\affiliation{School of Physical Sciences, University of Chinese Academy of Sciences,
Beijing 100049, China }

\author{Li Wang}
\email{liwangiphy@sxu.edu.cn}
\affiliation{Institute of Theoretical Physics, State Key Laboratory of Quantum Optics and Quantum Optics Devices, Collaborative Innovation Center of Extreme Optics, Shanxi University, Taiyuan 030006, P. R. China}

%\author{To be ap.}
\author{Shu Chen}
\email{schen@iphy.ac.cn }
%\thanks{schen@iphy.ac.cn }
\affiliation{Beijing National Laboratory for Condensed Matter Physics, Institute
of Physics, Chinese Academy of Sciences, Beijing 100190, China}
\affiliation{School of Physical Sciences, University of Chinese Academy of Sciences,
Beijing 100049, China }

\date{\today}
\begin{abstract}
We investigate the effect of an additional modulation parameter $\delta$ on the mobility properties of quasiperiodic lattices described by a generalized Ganeshan-Pixley-Das Sarma model with two on site modulation parameters. For the case with bounded quasiperiodic potential, we unveil the existence of self-duality relation, independent of $\delta$. By applying Avila's global theory, we analytically derive Lyapunov exponents in the whole parameter space, which enables us to determine mobility edges or anomalous mobility edges exactly. Our analytical results indicate that the mobility edge equation is described by two curves and their intersection with the spectrum gives the true mobility edge. Tuning the strength parameter $\delta$ can change the spectrum of the quasiperiodic lattice, and thus engineers the mobility of quasi-periodic systems, giving rise to completely extended, partially localized, and completely localized regions.  For the case with unbounded quasiperiodic potential, we also
obtain the analytical expression of the anomalous mobility edge, which separates localized states from critical states. By increasing the strength parameter $\delta$, we find
that the critical states can be destroyed gradually and finally
vanishes.
%The scheme proposed here could be implemented on the state of art experimental platforms, for example, the ultracold atoms in optical lattices.
\end{abstract}
\maketitle

\section{Introduction}
In condensed matter physics, mobility is a fundamental property of physical systems, which refers to the ability of a particle, such as an electron, to move through a material. In metal, it is responsible for the transport properties such as conductivity and resistance. In the context of semiconductors, it is an important parameter that determines the performance of electronic devices. Generally, mobility is influenced by factors like crystal structures, interactions, defects, and impurities, among others. More than sixty years ago, Anderson in his seminal work \cite{Anderson1958pr} investigated the role that the disordered on-site potential played on the mobility of particles in certain random lattices. From then on, Anderson localization \cite{Anderson1958pr,Abrahams1979prl} has attracted large and broad attention worldwide. Typically, for three-dimensional systems subjected to disorder of finite strength, localized and extended eigenstates can coexist in the energy band. Two intervals in the energy dimension corresponding to eigenstates with different mobility property are separated by a critical energy value, namely the mobility edge \cite{Mott1987jpc}. Tuning the strength of disorders may shift the value of mobility edge. Accordingly, the proportion between extended and localized eigenstates may also change, finally leading to the modulation of system's mobility.

While mobility edge is usually absent for the above-mentioned uncorrelated disorders in low dimensional systems \cite{Abrahams1979prl, RevModPhys.57.287},
%For example, in one-dimensional systems all eigenstates fall into completely localized even under infinitesimal small disorders.
one-dimensional (1D) quasiperiodic systems offer an appealing platform to study localization-delocalization transition \cite{AA1980,Thouless,Kohmoto,Kohmoto2008,Cai2013,Roati,Lahini} and mobility edge \cite{Ganeshan2015prl,Bloch,wyc20prl,mosaic-exp}. Among these, the most famous one is the Aubry-Andr\'{e} (AA) model \cite{AA1980}, which analytically demonstrates the existence of localization-delocalization transition by utilizing the self-duality property. Subsequently, various generalizations to the standard AA model confirmed the existence of mobility edge in 1D quasiperiodic lattices, for example, lattice models with slowly varying quasi-periodic potentials \cite{xiexc1988prl,xie90prb}, generalized AA model \cite{Ganeshan2015prl}, incommensurate lattices with exponentially decaying hoppings \cite{Biddle10prl}, and the recently proposed mosaic model \cite{wyc20prl}. So far,
the existence of mobility edges in low dimensional systems has been demonstrated in various models \cite{Hashimoto_1992, Boers07pra, PhysRevA.80.021603, PhysRevA.90.061602, Biddle11prb, Lixp16prb, Sarma17prb, Santos19prl, PhysRevB.100.174201, wyc20prl, Bloch, Roy21prl, PhysRevB.105.L081110, PhysRevLett.125.060401, SciPostPhys.12.1.027, SciPostPhys.13.3.046, PhysRevB.103.L060201,WangYC2021,ZhangYC,XuZH}. Very recently, the concept of mobility edge has found its new territory in the emerging field of non-Hermitian physics \cite{PhysRevB.106.024202,PhysRevB.105.054201,PhysRevB.105.054204,PhysRevB.106.144208,PhysRevB.101.174205,PhysRevB.103.214202,PhysRevB.100.125157,Tang_2021,LiuYX2021}.

In this work, we study quasiperidoic lattices described by a generalized Ganeshan-Pixley-Das Sarma (GPD) model with two tunable strength parameters of quasiperiodical potential. In comparison with the GPD model proposed by Ganeshan et. al \cite{Ganeshan2015prl}, also referred as generalized AA  model model in references, our model includes an additional modulation parameter $\delta$ (see Eq.(\ref{va})). By applying Avila's global theory, we analytically derive the Lyapunov exponent in the whole parameter space, which enables us to determine the mobility edge exactly. Our analytical results indicate that the mobility edge equation is independent of $\delta$ and generally  described by two curves, whose intersection with the spectrum of system gives the true mobility edges. Tuning the strength parameter $\delta$ can change the spectrum of the quasiperiodic lattice, and thus provides a scheme to engineer the mobility of quasi-periodic systems. In this manner, with the anchored mobility edge as a separation, the ratio of eigenstates on both sides then is changed, leading to the engineering of system's mobility.
Numerically calculating inverse participation ratios (IPRs) and Lyapunov exponents, we demonstrate that eigenstates of the system with bounded quasiperiodic potential successively cross the stationary mobility edge and undergo three scenarios, namely, completely extended, partially localized, and completely localized.
%Moreover, during the establishment of the proposal, we observe an interesting byproduct that the exact mobility edge formula implies that one can implement arbitrary modulation to the energy spectrum while keeping the mobility edges the same. Demonstrations of this kind of energy spectrum modulations are given in the last part of the main text.
For the case with unbounded quasiperiodic potential, we also obtain the analytical expression of the anomalous mobility edge, which separates localized states from critical states. By increasing the strength of $\delta$, we find that the critical states are destroyed gradually and finally vanish.

The paper is organized as follows.
First, we introduce our model in Sec. II A.
Subsequently, in Sec. II B, we unveil the existence self-duality relation for the system with bounded quasiperiodic potentials, independent of the modulation parameter $\delta$.  In  Sec. II C, by applying Avila's global theory, we derive analytically the expression of Lyapunov exponent and mobility edge.
In Sec. II D, we discuss the engineering of mobility and further verify our analytical results by numerically calculating the inverse participation ratios and Lyapunov exponents.
%Furthermore, we generalize the model to the case in which the two parameters are related to each other and more richer mobility properties are gotten.
The unbounded potential case is discussed in Sec. II E.
Finally, we give a summary in Sec. III.

\section{Model and results}
\subsection{model}
We consider a one-dimensional quasiperiodic lattice described by the following eigenvalue equation,
\begin{equation}
t \left ( \phi_{n-1}+\phi_{n+1} \right) + V_n(\lambda,\delta, \alpha) \phi_n =E \phi_n,
\label{vavbEqa}
\end{equation}
with
\begin{equation}
V_n(\lambda,\delta, \alpha) =\frac{\lambda \cos(2\pi n b+\theta)+\delta}{1-\alpha \cos(2\pi n b+\theta )}, \label{va}
\end{equation}
where $n$ is the index of lattice site, and $t$ is the nearest-neighbour hopping amplitude.
The quasiperiodic potential is regulated by two modulation parameters $\lambda$, $\delta$ and a deformation parameter $\alpha$. %which is constrained to the open interval $(-1,1)$.
The parameter $\theta$ denotes a phase factor and $b$ is an irrational number responsible for the quasiperiodicity of the on site potential. To be concrete, in this work we choose $b=(\sqrt{5}-1)/2$, however the obtained results are also valid for any other choice of the irrational number $b$.
For convenience, we shall set $t=1$ as the energy unit in the following calculation.

When $\delta=0$, the model reduces to the generalized AA model (GPD model) studied in ref.\cite{Ganeshan2015prl}, for which an exact mobility $\alpha E= 2 \text{sgn}(\lambda) |t|-\lambda$ is identified by the existence of a
generalized duality symmetry for the case of $\alpha \in (-1,1)$. On the other hand, the limit of $\lambda=0$ was recently studied in ref.\cite{SciPostPhys.12.1.027} for the unbounded case $\alpha>1$.  The onset of anomalous mobility edges at the energies $E = \pm 2 |t|$ is unveiled via the calculation of the Lyapunov exponent.

In this work, we shall consider the general case in the presence of both $\lambda$ and $\delta$ terms. For the bounded case with $\alpha \in (-1,1)$,  we unveil the existence of a self-dual symmetry even in the presence of $\delta$ term, which enables us to get an expression of mobility edge. By applying Avila's global theory, we can derive the mobility edges and  anomalous mobility edges analytically by calculating the Lyapunov exponents for both cases of  $|\alpha|<1$ and $|\alpha|>1$.

\subsection{Self-duality relation}
At first, we consider the case of $\alpha \in (-1,1)$ and demonstrate the existence of a generalized duality symmetry for the  model with the quasiperiodic potential (\ref{va}) under a
generalized dual transformation, from which we can derive the exact mobility edges by searching the self-duality relation. Following ref.\cite{Ganeshan2015prl}, we define
\[
\chi_n(\beta,\theta)\equiv \frac{\sinh{\beta}}{\cosh{\beta}-\cos(2\pi n b + \theta)}.
\]
Since Eq.(\ref{va}) can be represented as
\begin{equation}
V_n(\lambda,\delta, \alpha) =-\frac{\lambda}{\alpha} +\frac{\frac{\lambda}{\alpha} +\delta}{1-\alpha \cos(2\pi n b+\theta )}, \label{new potential}
\end{equation}
the model described by Eqs.~(\ref{vavbEqa}) and (\ref{va}) can be straightforwardly rewritten into a form as below,
\begin{equation}
t (\phi_{n-1}+\phi_{n+1}) + G\chi_n(\beta,\theta) \phi_n=\left( E+ \lambda \cosh{\beta}\right) \phi_n, \label{firstEq}
\end{equation}
in which $\beta$ is defined as $\cosh{\beta} \equiv 1/\alpha$ for $\alpha \in (0,1)$, and the parameter $G$ is given by $G=(\lambda \cosh{\beta}+\delta)\coth{\beta}$.

By using a well-established mathematical relation \cite{Ganeshan2015prl} as following,
\begin{equation}
\frac{\sinh{\beta}}{\cosh{\beta}-\cos(2\pi n b + \theta)}=\sum_{r=-\infty}^{\infty} e^{-\beta|r|} e^{i r (2\pi n b + \theta) }, \label{chip}
\end{equation}
we can implement consecutively three transformations to recover Eq.~(\ref{firstEq}) into its initial form. %At last, the analytical formula of mobility edge is obtained.
Define $u_p=\sum_n e^{in(2\pi bp+q\pi)}\phi_n$, where $\sum_n$ is short for $\sum_{n=-\infty}^{\infty}$ and $q$ is an integer. Multiplying $e^{in(2\pi bp+q\pi)}$ with both sides of Eq.~({\ref{firstEq}) and performing a summation, we get
\begin{equation}
\omega\chi_p^{-1}(\beta_0,0)e^{p\theta} u_p=G \sum_r e^{-\beta |p-r|} e^{r\theta} u_r,  \label{secondEq}
\end{equation}
where $\beta_0$ is defined through relation $E+\lambda \cosh{\beta} \equiv (-1)^q 2t \cosh{\beta_0}$ and $\omega$ is defined as $\omega\equiv (-1)^q 2t \sinh{\beta_0}$.
Subsequently, we move on to implement the second transformation $v_m=\sum_p e^{ip(2\pi bm+\theta+q\pi )}  \chi_p^{-1}(\beta_0,0) u_p $. By multiplying $e^{ip(2\pi bm+q\pi )}$ with both sides and making a sum over $p$, Eq.~(\ref{secondEq}) is correspondingly transformed into
\begin{equation}
\omega \chi_m^{-1}(\beta,q\pi) v_m=G\sum_r e^{-\beta_0 |m-r|} v_r. \label{thirdEq}
\end{equation}
Then it comes to the last step where the transformation is defined as $z_k=\sum_m e^{im(2\pi bk+\theta)} v_m$. We multiply Eq.~(\ref{thirdEq}) by $e^{im(2\pi bk+\theta)}$ and sum over $m$. Finally, one obtains the following tight binding model about $z_k$,
\begin{equation}
t(z_{k+1}+z_{k-1})+G\frac{\sinh{\beta}}{\sinh{\beta_0}} \chi_k(\beta_0, \theta) \;z_k =(-1)^q 2t \cosh{\beta}\;z_k. \label{finalEq}
\end{equation}
It is not difficult to notice that Eq.~(\ref{finalEq}) can be managed to be equivalent to Eq.~(\ref{firstEq}), if one lets $\beta=\beta_0$. Accordingly, we have $E+\lambda \cosh{\beta} = (-1)^q 2t \cosh{\beta}$, which in terms of the original parameter $\alpha$ is
\begin{equation}
\alpha E=(-1)^q 2t-\lambda. \label{exactME}
\end{equation}
Since $q$ may take even or odd integers, this actually gives out the analytical formula of a pair of exact mobility edges.
As for the other case $\alpha \in (-1,0)$, one can also arrive at Eq.~(\ref{exactME}) by conducting similar derivations as above. %Only minor adjustment is needed.

\subsection{Analytical formula of the exact mobility edge}

Next we apply Avila's global theory \cite{Avila} to calculate the Lyapunov exponent and derive the exact mobility edge \cite{LiuYX,LZC}.
For convenience, we will absorb $t$ into $\lambda$ and $E$ in the derivation process by setting $t=1$.

For the spectral problem with incommensurate potential, the Lyapunov exponent $\gamma(E)$ is defined as:
$$\gamma(E)=\lim_{L\to\infty}\frac{1}{ L} \ln||T_{L}(\theta)||,$$
where $||T_{L}(\theta)||$ is the norm of the $2\times 2$ transfer matrix $T_{L}(\theta)$, given by
\begin{equation}
T_L(\theta)=\prod_{n=1}^{L} M_n,
\end{equation}
in which
\begin{equation}
M_n = \begin{pmatrix}E-V_{n}&-1\\ 1&0\end{pmatrix},
\end{equation}
with $V_{n}$ given by Eq.(\ref{va}).

We adopt the conventional procedure to calculate Lyapunov exponent.
First, we need to  complex the phase, i.e., letting $\theta \rightarrow  \theta+i\epsilon$.
% with $\epsilon>0$.
In order to apply global theory more conveniently, we introduce a new matrix $\widetilde{M}_{j}$, which can be written as
\begin{equation}
\begin{split}
\widetilde{M}_{j}(\theta)&=[1-\alpha \cos(2\pi j b+\theta)]M_{j}.
\end{split}
\end{equation}
Then the transfer matrix for $\widetilde{M}_{j}(\theta)$ can be expressed as
$$\widetilde{T}_{L}(E,\theta)=\prod\limits_{j=1}^{L}\widetilde{M}_{j}(\theta).$$
And the Lyapunov exponent about $\widetilde{T}_{L}(E,\theta+i\epsilon)$ is
 $$\tilde{\gamma}(E,\theta+i\epsilon)=\lim_{L\to\infty}\frac{1}{L}  \ln||\widetilde{T}_{L}(E,\theta+i\epsilon)|| .$$
In the limit of $L\to \infty$, we can replace the sum of $j$ by an integral,
\begin{align*}
\tilde{\gamma}(E,\theta+i\epsilon)=\frac{1}{2\pi} \int  \ln||\widetilde{T}_{L}(E,\theta+i\epsilon)|| d\theta.
\end{align*}
Then it follows
\begin{align}\label{gamma1}
\gamma(E,\epsilon)&=\tilde{\gamma}(E,\epsilon)- \frac{1}{2\pi} \int \ln [1-\alpha \cos(\theta+i\epsilon) ]d \theta.
\end{align}
In this part, we focus on the case $-1<\alpha<1$ and the result of the integral in Eq.(\ref{gamma1}) is
$$
\frac{1}{2\pi} \int \ln (1-\alpha \cos(\theta+i\epsilon) )d \theta = \ln\frac{1+\sqrt{1- \alpha ^2}}{2},
$$
if $|\epsilon|<\ln|\frac{1+\sqrt{1- \alpha ^2}}{ \alpha }|$. From Eq.(\ref{gamma1}), we can find that $\gamma(E,\epsilon)$ and $\tilde{\gamma}(E,\epsilon)$ has the same slope about
$\epsilon$ when $|\epsilon|<\ln|\frac{1+\sqrt{1- \alpha ^2}}{ \alpha }|$.

In the large-$\epsilon$ limit, we get
\begin{equation} \label{T1}
\widetilde{T}_{L}(E,\epsilon)=\prod_{j=1}^L \frac{1}{2}e^{-i 2\pi b j}e^{|\epsilon|}\begin{pmatrix}- \alpha  E- \lambda& \alpha \\- \alpha &0\end{pmatrix}+o(1).
\end{equation}
According to the Avila's global theory, $\tilde{\gamma}(E,\epsilon)$ is a convex, piecewise linear function about $\epsilon\in(-\infty,\infty)$.
Combined with the result of Eq.(\ref{T1}), we can see that the slope about $\epsilon$ is always $1$.
%in the large-$\epsilon$ limit.
Thus, the Lyapunov exponent about $\widetilde{T}_{L}(E,\theta+i\epsilon)$ can be written as
%Besides, we can find the slope about $\epsilon$ is always $1$ in the large-$\epsilon$ limit.This means
$$\tilde{\gamma}(E,\epsilon)=|\epsilon|+\ln f(E),$$
for large enough $\epsilon$, where $$f(E)=|\frac{ |\alpha  E+\lambda|+\sqrt{( \alpha  E+\lambda)^2-4 \alpha ^2}}{4}|.$$

Considering the convexity of the Lyapunov exponent, the slope of $\gamma(E,\epsilon)$ might be 1 or 0 in the region $0\le |\epsilon|<\ln|\frac{1+\sqrt{1- \alpha ^2}}{ \alpha }|$.
%According to the Avila's global theory, Lyapunov eponent $\gamma(E,\epsilon)$ is convex.
%Thus the slope of $\gamma(E,\epsilon)$ might be 1 or 0 in the region $0\le \epsilon<\ln|\frac{1+\sqrt{1- \alpha ^2}}{ \alpha }|$.
Besides, the slope of $\gamma(E,\epsilon)$ in a neighborhood of $\epsilon=0$ is nonzero if the energy $E$ is in the spectrum.

Therefore, when $E$ is in the spectrum,
\begin{equation}
\tilde{\gamma}(E,\epsilon)=|\epsilon|+\ln f(E),
\end{equation}
for any $\epsilon\in(-\infty,\infty)$.
Based on Eq.(\ref{gamma1}) and the non-negativity of Lyapunov exponent $\gamma(E,\epsilon)$, we have
\begin{equation}\label{equation1.13}
\gamma(E,0)=\max\{\ln \frac{2f(E)}{1+\sqrt{1-\alpha^2}},0\},
\end{equation}
Then the mobility edge can be determined by $\gamma(E)=0$, which gives rise to
\begin{equation}
|\alpha \frac{E}{t} + \frac{\lambda}{t}|=2,
\label{ME2}
\end{equation}
where we have already explicitly included $t$.
\begin{figure}[bp]
\centering
\includegraphics[scale=0.45]{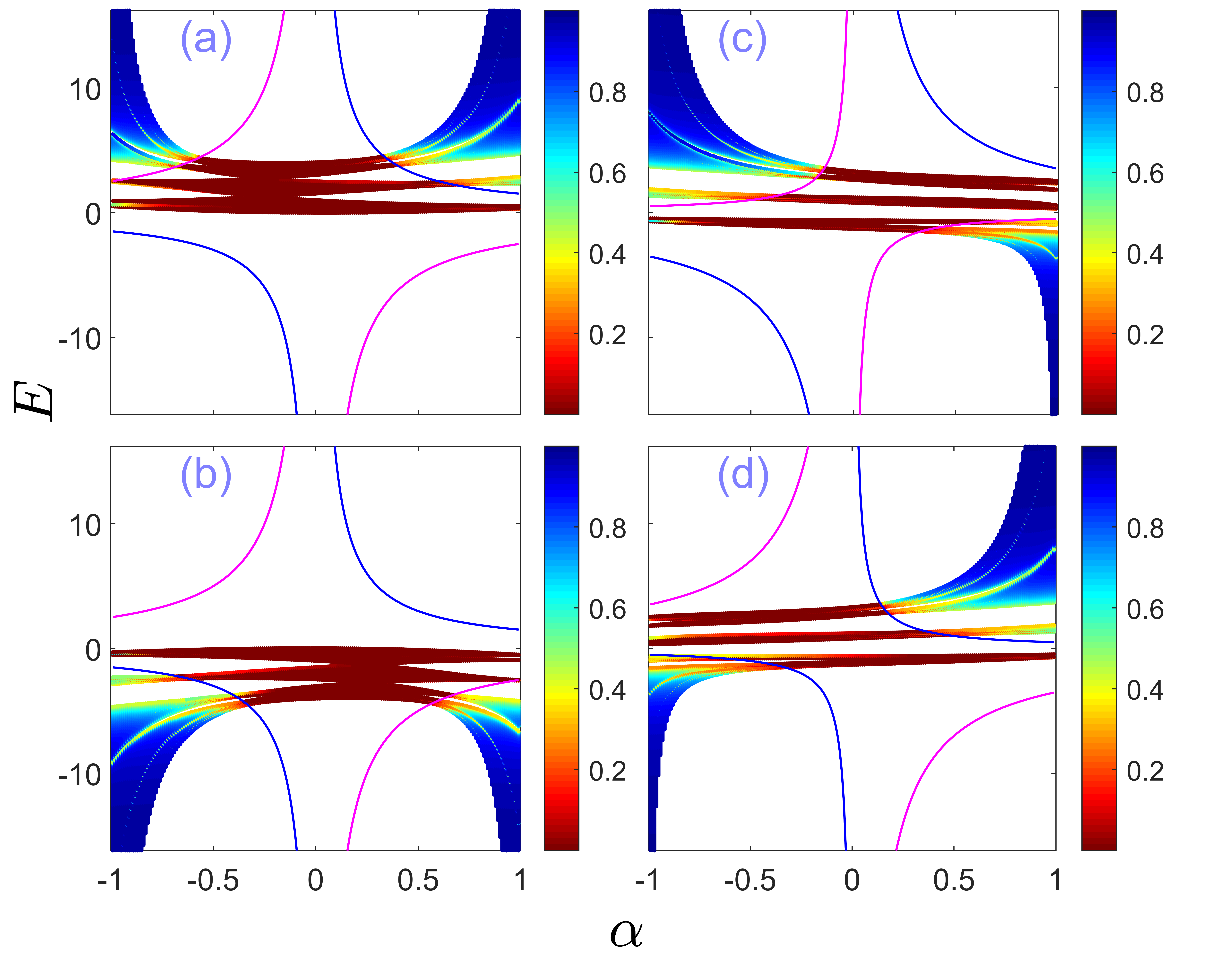}
\caption{Numerical spectrum $E$ of the model in Eq.~(\ref{vavbEqa}) as a function of $\alpha$ with model parameters $L=10000$, $\theta=0$ and $t=1$.
The IPR of each eigenstate is also calculated, which is indicated by the color of each eigenvalue in the spectrum.
 The lines in magenta and blue are exact mobility edges predicted by analytical formula Eq.~(\ref{exactME}).
(a) $\lambda=0.5$, $\delta=2$, (b) $\lambda=0.5$, $\delta=-2$, (c) $\lambda=-1.5$, $\delta=1$, (d) $\lambda=1.5$, $\delta=1$.}
\label{Fig02}
\end{figure}

Although Eq.(\ref{ME2}) takes a different form from Eq.(\ref{exactME}), it can be checked that they are actually equivalent.
This result suggests that the mobility edges may be composed of two curves.
The appearance of the mobility edge depends on another condition: a true mobility edge exists only if these curves are within the energy spectrum.
Therefore, the energy spectrum and the mobility edge equation together determine the mobility properties of the system.
In order to determine which curve determine the mobility edge for different parameters, we import the operator theory and give more accurate results.
By comparing the expression  of curves with the range of the physical possible energy spectrum (more details can be found in Appendix \ref{ME1}), we arrive at the expression:
\begin{equation}
E_c=\frac{2 \text{sgn}(\lambda+\delta \alpha)|t|-\lambda}{\alpha}. \label{MEE}
\end{equation}
When $\delta=0$, we see that the mobility edge reduces to $E_c=\frac{2 \text{sgn}(\lambda)|t|-\lambda}{\alpha}$, consistent with the result in Ref.\cite{Ganeshan2015prl}.
In this case, for a given $\lambda$ parameter, e.g., $\lambda>0$ and $t=1$, the mobility edge is only determined by the curve $E_c=\frac{2 - \lambda}{\alpha}$. However, in the presence of nonzero $\delta$, the mobility edge can be given by either $E_c=\frac{2 - \lambda}{\alpha}$ or $E_c=\frac{-2 - \lambda}{\alpha}$ depending on the value of $\lambda + \delta \alpha$, as displayed in Fig.\ref{Fig02}.

To gain an intuitive understanding, we display some numerical results in Fig.\ref{Fig02} for system with various parameters $\lambda$ and $\delta$, in which we display the energy spectrum versus $\alpha$ and plot the mobility edges given by Eq.(\ref{exactME}) and the inverse participation ratios (IPRs)\cite{THOULESS197493} as a function of $\alpha$.
The IPR for an eigenstate with eigenvalue $E$ is given as
\begin{equation}
\mathrm{IPR}(E_i)=\frac{\sum_n |\phi_n(E_i)|^4}{\left(\sum_n |\phi_n(E_i)|^2\right)^2}, \label{IPRE}
\end{equation}
where $E_i$ is the $i$-th energy eigenvalue.
For an extended eigenstate, the probability tends to be distributed evenly among the lattice, thus the IPR is expected to be the order of $1/L$. While for a localized eigenstate, the probability is usually well confined to a few lattice sites, therefore the IPR approaches $1$ in the limiting case. It is shown that the localized and extended region are separated by the mobility edge.
In Fig.\ref{Fig02}(a-b), the mobility edges are determined by different curves because the sign of $\lambda+\delta \alpha $ is changed in the process of adjusting $\alpha$ from $-1$ to $ 1$.
In contrast, the mobility edges in Fig.\ref{Fig02}(c-d) are determined by only one curve because adjusting $\alpha$ does not change the sign of $\lambda+\delta \alpha$.

\subsection{Engineering the mobility property}

Although the equation of mobility edges are simply two straight lines described by $E=\frac{2}{\alpha}-\frac{\lambda}{\alpha}$ and $E=-\frac{2}{\alpha}-\frac{\lambda}{\alpha}$, which are independent of $\delta$, tuning $\delta$ can change the spectrum of the system dramatically.
By tuning $\delta$, we can access five different regions as shown in Fig.\ref{ChangeDelta}.

By comparing the energy spectrum and the equation of mobility edges, we can approximately obtain transition points separating these different regions of $\delta$ (details about the transition points can be found in Appendix \ref{transition point}).
For the case of $-1<\alpha<0$, as shown in Fig.\ref{ChangeDelta}(a), the five different regions are:
(i) For $-\infty<\delta< -\frac{\lambda}{\alpha}+\frac{2(1-\alpha)}{\alpha}$, all the eigenstates are localized;
(ii) For $ -\frac{\lambda}{\alpha}+\frac{2(1-\alpha)}{\alpha} <\delta<-\frac{\lambda}{\alpha}+\frac{2(1+\alpha)^2}{\alpha} $, there is a mobility edge determined by $E=-\frac{\lambda}{\alpha}+\frac{2}{\alpha}$, below which the states are localized, whereas above which the states are extended;
(iii) For $ -\frac{\lambda}{\alpha}+\frac{2(1+\alpha)^2}{\alpha}<\delta< -\frac{\lambda}{\alpha}-\frac{2(1+\alpha)^2}{\alpha}$, all the eigenstates are extended;
(iv) For $-\frac{\lambda}{\alpha}-\frac{2(1+\alpha)^2}{\alpha}<\delta< -\frac{\lambda}{\alpha}-\frac{2(1-\alpha)}{\alpha}$,  there is a mobility edge determined by $E=-\frac{\lambda}{\alpha}-\frac{2}{\alpha}$, below which the states are extended, whereas above which the states are localized;
(v) For $-\frac{\lambda}{\alpha}-\frac{2(1-\alpha)}{\alpha}<\delta <+ \infty $, all the eigenstates are localized.
For the case of $0<\alpha<1$, as shown in Fig.\ref{ChangeDelta}(b), the five different regions are:
(i) For $-\infty<\delta< -\frac{\lambda}{\alpha}-\frac{2(1+\alpha)}{\alpha}$, all the eigenstates are localized;
(ii) For $ -\frac{\lambda}{\alpha}-\frac{2(1+\alpha)}{\alpha} <\delta<-\frac{\lambda}{\alpha}-\frac{2(1-\alpha)^2}{\alpha} $, there is a mobility edge determined by $E=-\frac{\lambda}{\alpha}-\frac{2}{\alpha}$, below which the states are localized, whereas above which the states are extended;
(iii) For $ -\frac{\lambda}{\alpha}-\frac{2(1-\alpha)^2}{\alpha}<\delta< -\frac{\lambda}{\alpha}+\frac{2(1-\alpha)^2}{\alpha}$, all the eigenstates are extended;
(iv) For $-\frac{\lambda}{\alpha}+\frac{2(1-\alpha)^2}{\alpha}<\delta< -\frac{\lambda}{\alpha}+\frac{2(1+\alpha)}{\alpha}$,  there is a mobility edge determined by $E=-\frac{\lambda}{\alpha}+\frac{2}{\alpha}$, below which  states are extended, whereas above which states are localized;
(v) For $-\frac{\lambda}{\alpha}+\frac{2(1+\alpha)}{\alpha}<\delta <+ \infty $, all the eigenstates are localized.

\begin{figure}[bp]
\centering
\includegraphics[scale=0.2217]{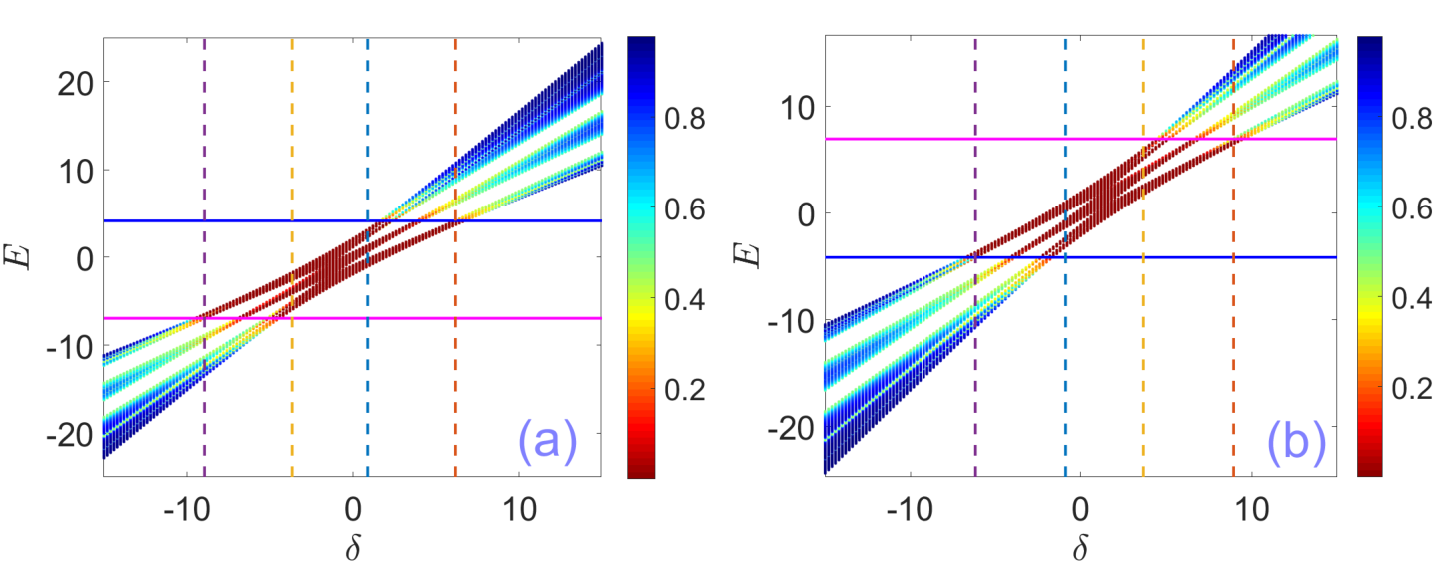}
\caption{Numerical spectrum $E$ of the model in Eq.~(\ref{vavbEqa}) as a function of $\delta$ with different parameters.
(a)$\lambda=-0.5$, $\alpha=-0.36$.
(b)$\lambda=-0.5$, $\alpha=0.36$.
We choose $L=10000$, $\theta=0$ and $t=1$ in all cases.
The lines in magenta and blue are exact mobility edges predicted by analytical formula Eq.~(\ref{exactME}).
The dashed lines denote transition points separating different regions.
\label{ChangeDelta}}
\end{figure}

%\begin{figure*}[t]    %  span over two columns
\begin{figure}[btp]
\centering
\includegraphics[scale=0.4]{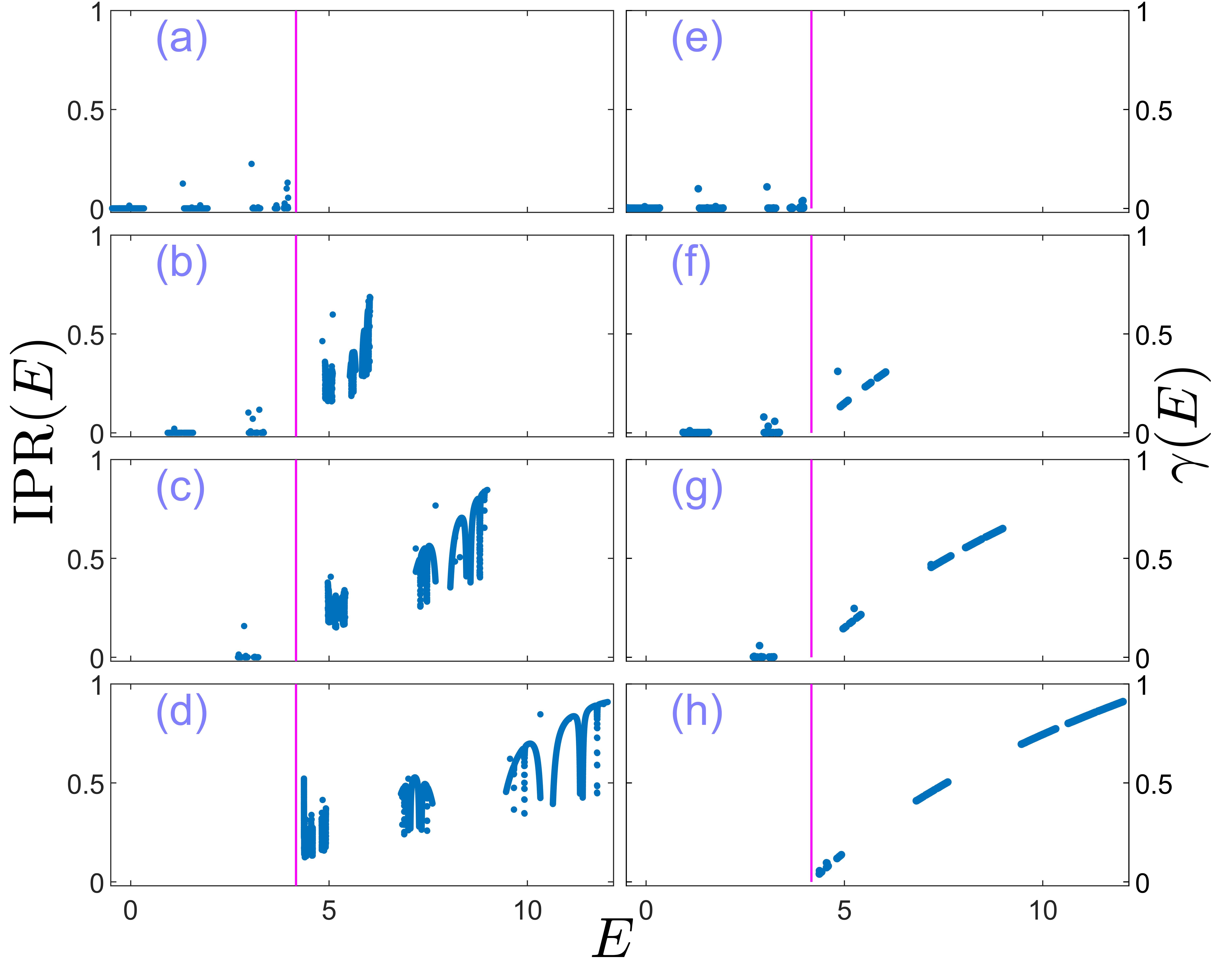}
\caption{Engineering the system's mobility by varying strength $\delta$ of the second quasi-periodic on-site potential while mobility edge is kept fixed by the first quasi-periodic on-site potential strength $\lambda$. The left column shows the inverse participation ratios (IPRs) of all single-particle eigenstates for different values of $\delta$. The lattice size is $L=10000$ with parameters $t=1$, $\theta=0$, $\alpha=-0.36$, and $\lambda=-0.5$. The right column gives the corresponding Lyapunov exponents (LEs).
(a,e) $\delta=1.5$, (b,f) $\delta=3.0$, (c,g) $\delta=5.0$, (d,h) $\delta=7.0$. The vertical line in (a-h) denotes position of the anchored mobility edge.
}
\label{Fig01}
\end{figure}
%\end{figure*}

%We numerically diagonalize the tight binding model described by Eq.~(\ref{vavbEqa}) for lattice with finite size $L$ to verify our results.
%To display the procedure how the system's mobility is engineered by varying $\delta$ whilst still keeping the mobility edge unchanged, we calculate IPRs to characterize the localization properties of single-particle eigenstates.

To see how the mobility is engineered by the strength of $\delta$,  we show the change of IPRs and Lyapunov exponents of all eigenstates in Fig.~\ref{Fig01} by choosing several typical parameters corresponding to Fig.2(a). The Lyapunov exponents \cite{xie90prb} (LEs) for finite-size lattices can be numerically calculated by using \cite{Thouless72,Lix20prb}
\begin{equation}
\gamma(E_i)=\frac{1}{L-1}\sum_{j \ne i} \ln \left|  \frac{E_i-E_j}{t} \right|. \label{LE}
\end{equation}
It is well-known that Lyapunov exponent is the inverse of localization length, thus for an extended eigenstate it approaches to a vanishing value as the lattice size $L$ increases. On the other hand, the Lyapunov exponent is non-zero for localized states. The IPRs for all single-particle eigengstates under different strengths of $\delta$ are shown in Fig.~\ref{Fig01}(a-d) and the LEs are correspondingly given in Fig.~\ref{Fig01}(e-h). For all of them the strength of $\lambda$ is fixed at $\lambda=-0.5$. The lattice size is $L=10000$ and other parameters are $\alpha=-0.36$ and $\theta=0$. From top to bottom, the corresponding strengths of the second quasi-periodic potential are $\delta=1.5$, $\delta=3.0$, $\delta=5.0$, and $\delta=7.0$.
It is clearly shown that as the strength of $\delta$ is modulated from $\delta=1.5$ to $\delta=7.0$, the system is engineered to undergo different situations, initially wholly extended, then partially localized, and at last completely localized. Notably, during the whole process, the mobility edge denoted by vertical line in Fig.~\ref{Fig01} is fixed and rather robust against the variation of the strength of $\delta$. As the strength of $\delta$  is varied, single particle eigenstates change their mobility properties by leapfrogging the fixed mobility edge consecutively, one by one.

%It is worth mentioning that the alternative manner of mobility engineering is more friendly to experiments in that in traditional method when one tune the relevant parameters of the system, both energy spectrum and mobility edge change simultaneously. In that case, one has to measure and monitor the spectrum and mobility edge at the same time to engineer the system's mobility.

\begin{figure}[tbp]
\centering
\includegraphics[scale=0.26]{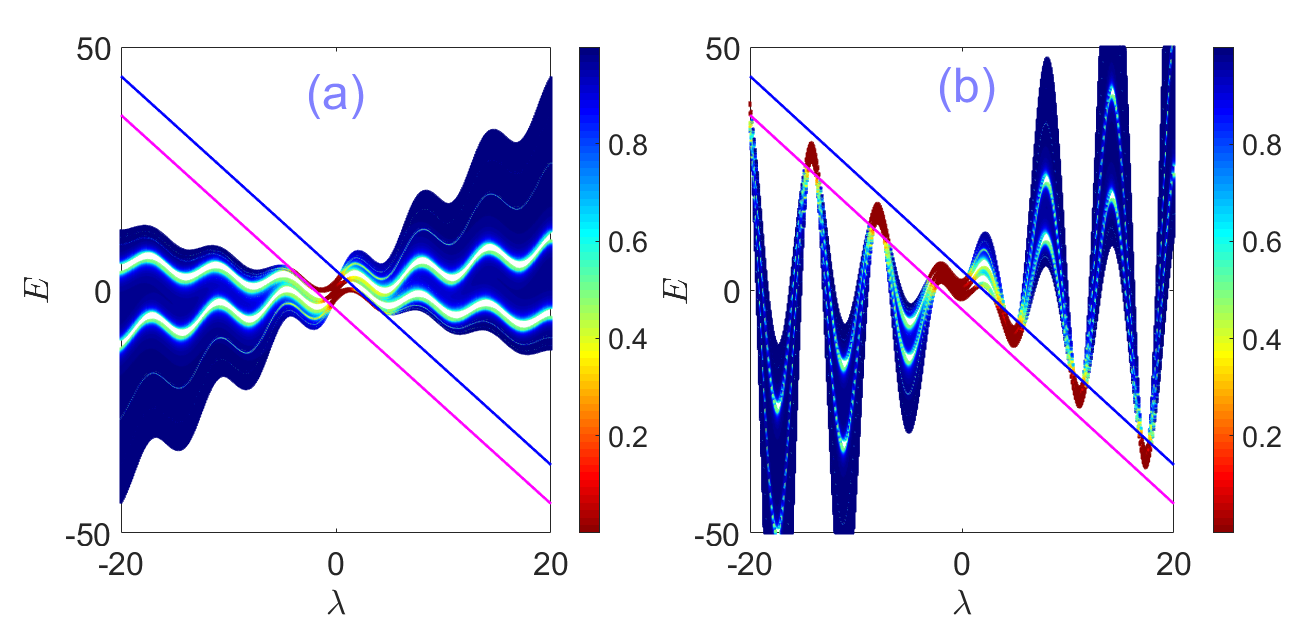}
\caption{Examples of energy spectrum engineering with the freedom granted by the second quasi-periodic potential. IPR is indicated by the color of the eigenvalue point. The lattice size is $L=10000$.
(a) $\delta=\frac{1}{\alpha} \sin{(\lambda)}$, (b) $\delta=\frac{\lambda}{\alpha}\sin(\lambda)$.
Other parameters are $\alpha=0.5$, $t=1$, $\theta=0$.
}
\label{tiaozhi}
\end{figure}

In the above calculation, $\delta$ is chosen as an independent parameter. Nevertheless, we can also choose $\delta$ as a function of $\lambda$. Although the form of $\delta(\lambda)$ does not change the mobility edge equation, it can modulate the structure of spectrum and thus enable us engineering the mobility properties of the quasiperiodic lattices.  In Fig.~\ref{tiaozhi}(a) and Fig.\ref{tiaozhi}(b), we display the energy spectrum and corresponding IPRs versus $\lambda$ for systems with $\delta=\frac{1}{\alpha} \sin(\lambda)$ and $
\delta=\frac{\lambda}{\alpha} \sin(\lambda)$, respectively.
While the extended states and the mobility edges occur only in a region around $\lambda=0$ as shown in Fig.~\ref{tiaozhi}(a), we find that the mobility edges occur periodically in Fig.~\ref{tiaozhi}(b) with the increase of $\lambda$.
Intuitively, periodically occurring mobility edges can be attributed to the periodical occurrence of zero points of $\frac{\lambda}{\alpha}+\delta(\lambda)$.
According to the expression of Eq.(\ref{new potential}), when $\frac{\lambda}{\alpha}+\delta(\lambda)=0$, the quasiperiodic potential vanishes, and the corresponding eigenstates must be extended states. When $\frac{\lambda}{\alpha}+\delta(\lambda) \neq 0$, localized states may occur if the energy spectrum exceeds the mobility edge curves.

\subsection{Anomalous mobility edges for the case of $|\alpha|>1$}

\begin{figure}[btp]
\centering
\includegraphics[scale=0.187]{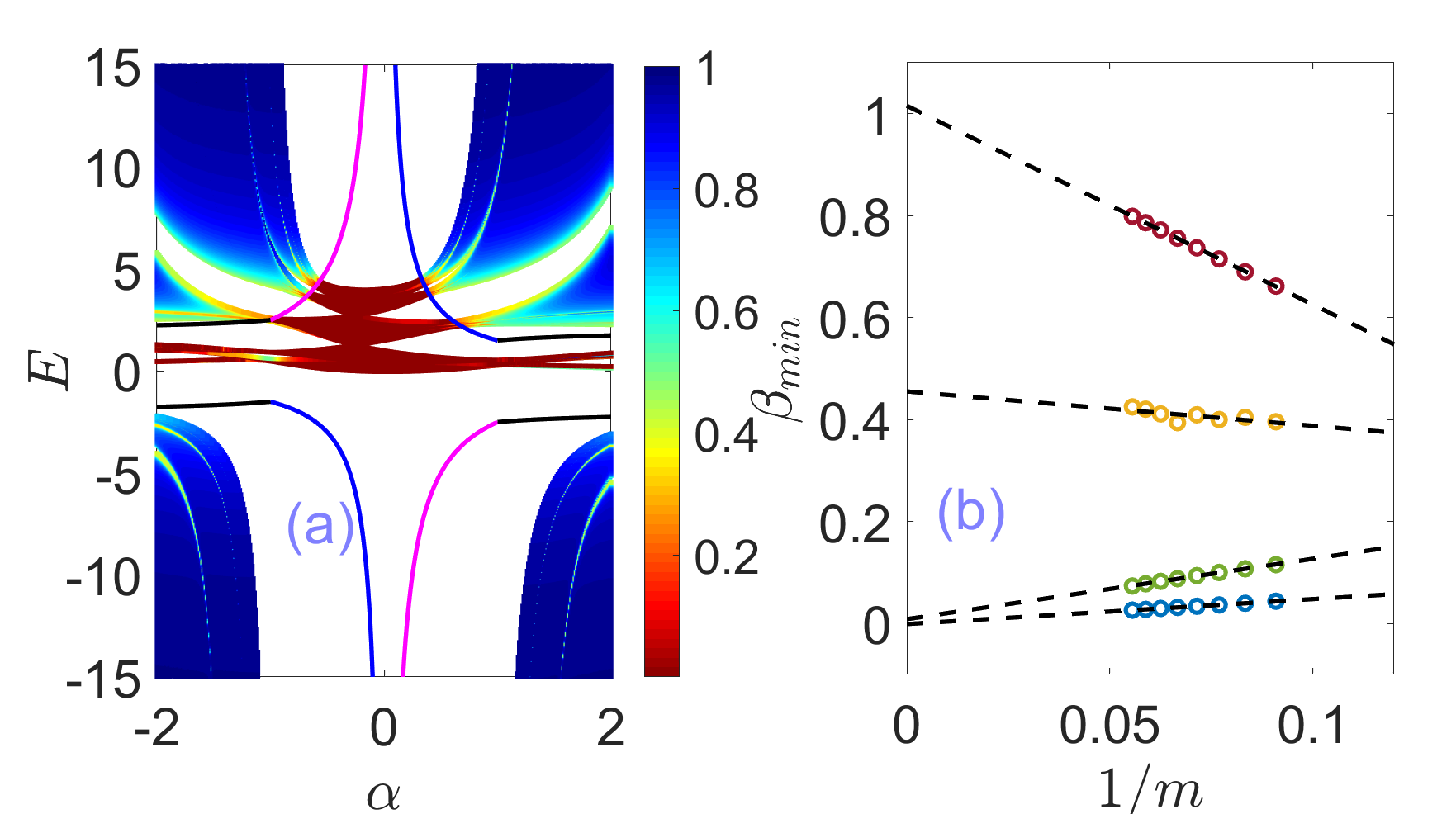}
\caption{
(a)Mobility edges and anomalous mobility edges.
The lattice size is $L=10000$.
Other parameters are $t=1$, $\theta=0$, $\delta=2$ and $\lambda=0.5$.
The lines in magenta and blue are exact mobility edges predicted by analytical formula Eq.~(\ref{exactME}).
The lines in black are anomalous mobility edges predicted by analytical formula Eq.(\ref{MEE2}).
(b) $\beta_{min}$ as a function of the inverse Fibonacci index $1/m$ for different $\alpha$.
From top to bottom, the data points with different color represents,
extended eigenstates (in the energy interval $[\frac{2}{\alpha}-\frac{\lambda}{\alpha},-\frac{2}{\alpha}-\frac{\lambda}{\alpha}]$) with $\alpha=-0.36$,
critical eigenstates (in the energy interval $[-2-\frac{\lambda}{\alpha},2-\frac{\lambda}{\alpha}]$)  with $\alpha=1.5$,
localized eigenstates (outside the energy interval $[\frac{2}{\alpha}-\frac{\lambda}{\alpha},-\frac{2}{\alpha}-\frac{\lambda}{\alpha}]$) with $\alpha=-0.36$,
localized eigenstates (outside the energy interval $[-2-\frac{\lambda}{\alpha},2-\frac{\lambda}{\alpha}]$)  with $\alpha=1.5$.
We choose $\lambda=0.5$ ,$\theta=0$ and $\delta=0.5$ in our calculation.
}
\label{alphaL1}
\end{figure}

For the case of  $|\alpha|>1$, the quasiperiodic potential given by Eq.(\ref{va}) is in principle an unbounded potential, which, however, does not diverge
at any lattice site for a finite size lattice. According to the Simon-Spencer theorem \cite{Simon}, extended states are forbidden for an unbounded quasiperiodic potential, and thus the self-duality mapping does not work. Nevertheless, we can use Avila's global theory for unbounded quasiperiodic operators to derive the analytical expression of anomalous mobility edges \cite{ZhangYC,SciPostPhys.12.1.027}.
The derivation of mobility edges for $|\alpha|>1$ is similar to the case of $|\alpha|<1$ until Eq.(\ref{gamma1}).
The result of the intergal in Eq.(\ref{gamma1}) for $|\alpha|>1$ is
$$
\frac{1}{2\pi} \int \ln [1-\alpha \cos(\theta+i\epsilon) ]d \theta = |\epsilon|+\ln(\frac{\alpha}{2}).
$$
Thus we can get the Lyapunov exponent in the large-$\epsilon$ limit as
\begin{align*}
\gamma(E,\epsilon)&=\ln(\frac{2f(E)}{\alpha})
\end{align*}
for any $\epsilon$.
The Lyapunov exponent $\gamma(E,\epsilon)$ is independent of $\epsilon$.
Similar to the discussion in ref.\cite{SciPostPhys.12.1.027}, there is an anomalous mobility edge determined by $\gamma(E)=0$.
Here the anomalous mobility edge means an edge separating localized states and critical states.
Through straighforward calculations, we arrive at an exact analytical formula of the anomalous mobility edge as
\begin{align}\label{MEE2}
E_c=\pm 2 |t| -\frac{\lambda}{\alpha}.
\end{align}
Before proceeding further discussion, we set $t=1$ for convenience.
In regions of $E>2-\frac{\lambda}{\alpha}$ and $ E<-2-\frac{\lambda}{\alpha}$,  $\gamma(E)>0$ and the eigenstates are localized eigenstates with localization length $\xi=1/\gamma(E)$.
In the region $-2-\frac{\lambda}{\alpha}<E<2-\frac{\lambda}{\alpha}$, the energy spectrum is singular continuous and the eigenstates are critical.

\begin{figure}[btp]
\centering
\includegraphics[scale=0.37]{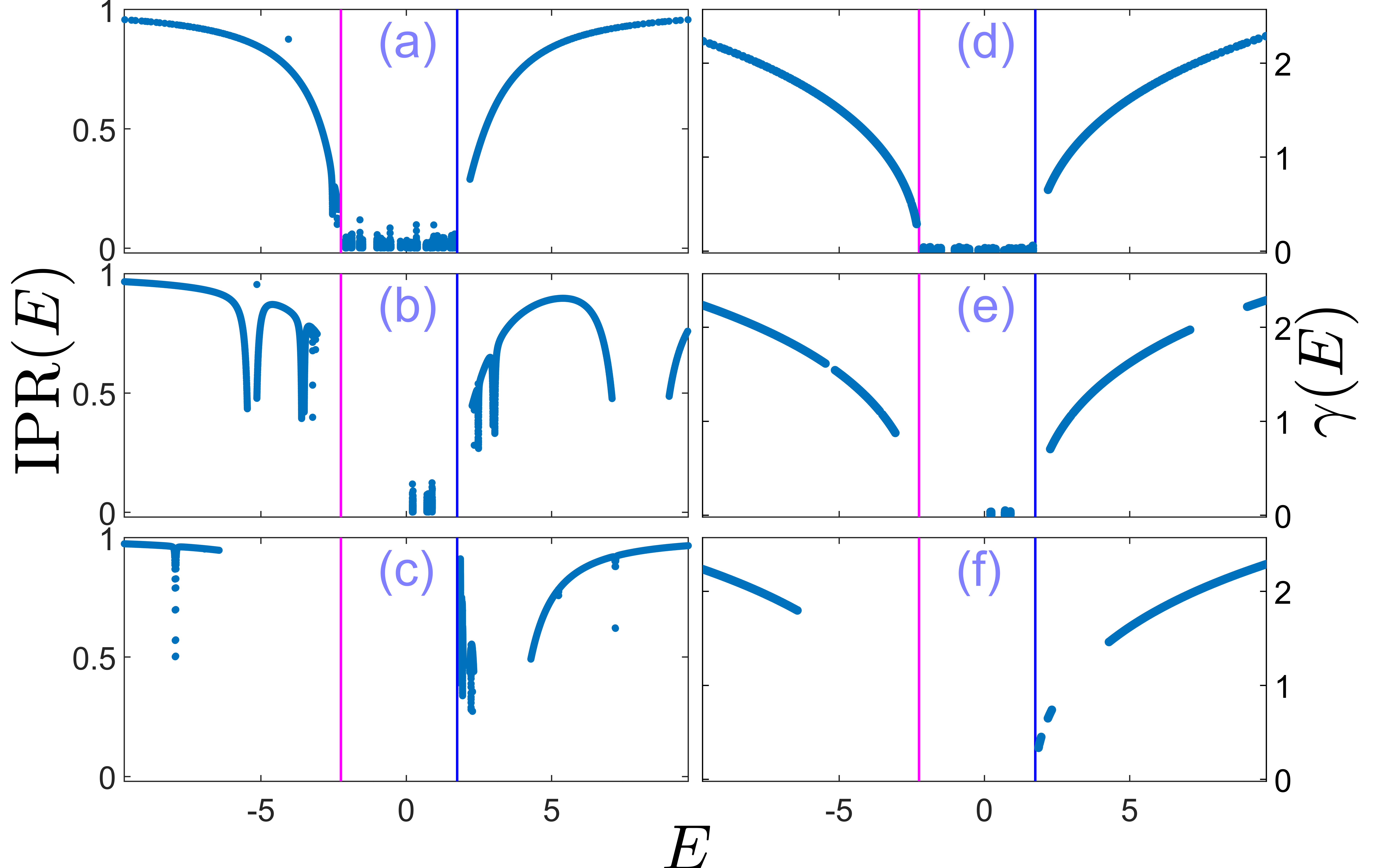}
\caption{Modulations of eigenstate properties by varying strength $\delta$ of the second quasi-periodic on-site potential while mobility edge is kept fixed by the first quasi-periodic on-site potential strength $\lambda$. The left column shows the inverse participation ratios (IPRs) of all single-particle eigenstates for different values of $\delta$. The lattice size is $L=10000$ with parameters $t=1$, $\theta=0$, $\alpha=-2$, and $\lambda=-0.5$. The right column gives the corresponding Lyapunov exponents (LEs).
(a,d) $\delta=0$, (b,e) $\delta=2.0$, (c,f) $\delta=5.7$. The vertical lines in (a-f) denote positions of the anchored anomalous mobility edges given by Eq.(\ref{MEE2}).
}
\label{Fig06}
\end{figure}

Next we carry out numerical analysis to unveil the existence of anomalous mobility edges in the regime of $|\alpha |>1$. In Fig.\ref{alphaL1}(a),  we display the energy spectrum and corresponding IPRs versus $\alpha$ for both the regions of $|\alpha |<1$ and $|\alpha |>1$.
In order to distinguish the extended eigenstates and critical eigenstates displayed in Fig.\ref{alphaL1}(a),
we make multifractal analysis and calculate the scaling exponent $\beta_{min}$.
The multifractal analysis demands considering a series of finite systems with different sizes. We thus choose the system size $L$ as the $m$th Fibonacci number $F_m$.
The scaling exponent $\beta_{min}$ can be extracted as follows.
For a given wave function $\psi_n^j$, one can extract a scaling exponent $\beta_n^j$  from the $n$th on-site probability,
$
P_n^j=|\psi_n^j|^2 ~(1/F_m)^{\beta_n^j}
$.
Here we use the minimum value
$
\beta_{\min}^j=\min_n(\beta_n^j)
$
to characterize eigenstate properties.
As the system size increases, $\beta_{\min}^j\rightarrow 1$ for the extended eigenstates, whereas
$\beta_{\min}^j\rightarrow 0$ for the localized eigenstates.
For the critical eigenstates, the $\beta_{\min}^j$ approches to a value in the interval $(0,1)$.
In order to reduce the fluctuations among different critical eigenstates, we define an average scaling exponent
$
\beta_{\min}=\frac{1}{L'}\sum_{j=1}^{L'} \beta_{\min}^j,
$
where $L'$ is the number of eigenstates in the corresponding region.
In Fig.\ref{alphaL1}(b), the numerical result of scaling analysis is shown. For the regime of $|\alpha|>1$, there appear anomalous mobility edges.
On the other hand, there are normal mobility edges for the regime of $|\alpha|<1$.

From Eq.(\ref{MEE2}), we see that the pair of anomalous mobility edges is completely independent of $\delta$.
Thus in unbounded case, one is also granted a degree of freedom to engineer the system's spectrum while the position of the anomalous mobility edge is kept fixed. As the strength of $\delta$ varies, certain eigenstate may hop across the anomalous mobility edge and the property of the eigenstate changes. In Fig. \ref{Fig06}, we show this manner of modulations of eigenstate properties by numerically calculating IPRs (left column) and LEs (right coloumn) for all eigenstates. The two vertical lines denote the anomalous mobility edges predicted by Eq.(\ref{MEE2}). Data points in between stand for critical eigenstates while those points outside denote localized eigenstates. From top to bottom, the strength of the second quasiperiodic potential are $\delta=0, 2$ and $5.7$. It is clearly shown that as $\delta$ varies, the critical states are killed gradually and finally all critical states vanish. For the unbounded case of $|\alpha|>1$, we notice that the spectrum is very wide and thus a region with all eigenstates being critical states is hard to be accessed  by  tuning $\delta$, which is in contrast with the bounded case where a completely extended region is accessible.

\section{Summary}
In summary, we study 1D quasiperiodic lattices described by a generalized GPD  model with an additional tunable parameter $\delta$ in the whole parameter space, including cases with both the bounded and unbounded quasiperiodic potential.
By applying Avila's global theory, we derive the analytical expression of Lyapunov exponent, which permits us to get the exact expression of mobility edges and anomalous mobility edges. Although the mobility edge equation and anomalous mobility edge equation do not include the introduced parameter $\delta$ explicitly,
the parameter can modulate the energy spectrum and thus provides a way to engineering the mobility properties of the system. By numerically calculating the IPRs and Lyapunov exponents,
we show that the mobility can be flexibly engineered by modulating the strength of new parameter while the mobility edge equation is kept unchanged. For the bounded case, the modulation of $\delta$ can lead to completely extended, partially localized, and completely localized regions. For the unbounded case, the modulation of $\delta$ can only lead to partially localized and completely localized states, whereas a completely critical region is hard to be accessed.
Our study unveils the richness of quasiperiodic localization and provides a scheme to engineer the mobility properties of quasiperiodic lattices.

%Given the rapid development of the experimental side, these findings could be hopefully verified experimentally in synthetic lattices of laser-coupled momentum modes along the lines of works \cite{Ma22prl,Gadway15pra,Gadway16pra,An21prl} and other state of art platforms.

\begin{acknowledgments}
 L.W. is supported by the Fundamental Research Program of Shanxi Province, China (Grant No. 202203021211315), the National Natural Science Foundation of China (Grant Nos. 11404199, 12147215) and the Fundamental Research Program of Shanxi Province, China (Grant Nos. 1331KSC and 2015021012). S. C. is supported by the NSFC under Grants No. 12174436 and
No. T2121001 and the Strategic Priority Research Program of Chinese Academy of Sciences under Grant No. XDB33000000.
\end{acknowledgments}

\appendix
\section{Accurate expression of the model's mobility edge for the case with $|\alpha|<1$ \label{ME1}}

The mobility edge can be determined by letting Lyapunov exponent $\gamma(E)=0$, which gives
\begin{align}
\left|\frac{\alpha E+\lambda}{t}\right|=2. \label{MEpaireq}
\end{align}
To be specific, it consists of two parts,
\begin{align}
E_{c1}&=\frac{2|t|-\lambda}{\alpha}, \label{MEa} \\
E_{c2}&=\frac{-2|t|-\lambda}{\alpha}. \label{MEb}
\end{align}

To get a more accurate formula for the mobility edge, one has to resort to operator theory.

According to the operator theory, the range of the physical possible energy spectrum $E$ of the model Eq.(\ref{vavbEqa}) can be estimated as ${E}\subseteq [-2|t|+\min(V_n),2|t|+\max(V_n)]$.

Before proceeding, we note that the on site potential can be rewritten as
\begin{equation}
V_n=\frac{\lambda/\alpha+\delta}{1-\alpha\cos(2\pi nb+\theta)}-\lambda/\alpha.
\end{equation}
Thus when $\lambda/\alpha+\delta\textgreater0$ and $\alpha\textgreater0$, we have
\begin{equation}
\{E\}\subseteq[-2|t|+\frac{\lambda/\alpha+\delta}{1+\alpha}-\lambda/\alpha,2|t|
+\frac{\lambda/\alpha+\delta}{1-\alpha}-\lambda/\alpha],
\label{4B}
\end{equation}
while when $\lambda/\alpha+\delta\textgreater0$ and $\alpha\textless0$, we have
\begin{equation} \{E\}\subseteq[-2|t|+\frac{\lambda/\alpha+\delta}{1-\alpha}-\lambda/\alpha,
2|t|+\frac{\lambda/\alpha+\delta}{1+\alpha}-\lambda/\alpha].
\label{4B}
\end{equation}
And when $\lambda/\alpha+\delta\textless0$ and $\alpha\textgreater0$, we have
\begin{equation}	\{E\}\subseteq[-2|t|+\frac{\lambda/\alpha+\delta}{1-\alpha}-\lambda/\alpha,
2|t|+\frac{\lambda/\alpha+\delta}{1+\alpha}-\lambda/\alpha],
\label{4B}
\end{equation}
while $\lambda/\alpha+\delta\textless0$ and $\alpha\textless0$, we have
\begin{equation} \{E\}\subseteq[-2|t|+\frac{\lambda/\alpha+\delta}{1+\alpha}-\lambda/\alpha,
2|t|+\frac{\lambda/\alpha+\delta}{1-\alpha}-\lambda/\alpha].
	\label{4B}
\end{equation}

According to the above-obtained ranges of the energy spectrum $E$ under four different cases, we can arrive at
more accurate mobility edges by excluding the un-physical part.

Firstly, we consider the case with $\lambda/\alpha+\delta\textgreater0$ and $\alpha\textgreater0$. In this case, it is obviously that $E_{c1}>E_{c2}$.  And we have the following relation,
\begin{equation}
-2|t|\alpha+\frac{\lambda+\delta\alpha}{1+\alpha}\textgreater-2|t|.
\end{equation}
So accordingly one can get
\begin{equation}    	
E_{c2}< -2|t|+\frac{\lambda/\alpha+\delta}{1+\alpha}-\lambda/\alpha
\end{equation}
This means that $E_{c2}$ is even below the lower limit of the energy spectrum.
So $E_{c2}$ should be omitted and only $E_{c1}$ is valid in this case.

Secondly, we turn to the case $\lambda/\alpha+\delta\textgreater0$ and $\alpha\textless0$, for which
we have $E_{c2}>E_{c1}$. Noting that $\lambda+\delta \alpha <0$, it is easily to find that the following relation is fulfilled,
\begin{equation}
2|t|(1+\alpha)(1-\alpha)\textgreater\lambda+\delta\alpha.
\end{equation}
Thus we can see that $E_{c1}$ is lower than the minimum of the model's energy spectrum $E$, i.e.,
\begin{equation}
E_{c1} < -2|t|+\frac{\lambda/\alpha+\delta}{1-\alpha}-\lambda/\alpha.
\end{equation}
So in this case, $E_{c1}$ is excluded and $E_{c2}$ is kept.

Thirdly, we consider the case $\lambda/\alpha+\delta\textless0$ and $\alpha\textgreater0$.
In this case, we have $E_{c1}>E_{c2}$ and relation,
\begin{equation}
\frac{\lambda/\alpha+\delta}{1+\alpha}\textless 0\textless 2|t|(\frac{1}{\alpha}-1).
\end{equation}
It is straightforward to arrive at,
\begin{equation}
E_{c1}> 2|t|+\frac{\lambda/\alpha+\delta}{1+\alpha}-\lambda/\alpha,
\end{equation}
which means $E_{c1}$ is outside the range of the model's energy spectrum.
Therefore, in this case, the model's mobility edge is determined by $E_{c2}$.

Fourthly, we check the case $\lambda/\alpha+\delta\textless0$ and $\alpha < 0$.
Obviously, we have $E_{c2}>E_{c1}$ in this case. Also noting the following realtion
\begin{equation}
\frac{\lambda/\alpha+\delta}{1-\alpha}\textless0\textless -2|t| (\frac{1}{\alpha}+1),
\end{equation}
we can get
\begin{equation}
E_{c2} > 2|t|+\frac{\lambda/\alpha+\delta}{1-\alpha}-\lambda/\alpha.
\end{equation}
This means $E_{c2}$ is above the upper limit of the physical model's energy spectrum $E$.
Therefore, the mobility edge in this case is determined by $E_{c1}$.

In summary, when $\alpha\textgreater0$, the mobility edge can be described by
\begin{equation}
E_c=\frac{2\text{sgn}(\lambda/\alpha+\delta)|t|-\lambda}{\alpha},
\end{equation}
and on the other hand, for $\alpha < 0$, we have
\begin{equation}
E_c=\frac{-2\text{sgn}(\lambda/\alpha+\delta)|t|-\lambda}{\alpha}.
\end{equation}

Furthermore, the mobility edge can be written in a briefer form,
\begin{equation}
	E_c=\frac{2\sgn(\alpha)\sgn(\lambda/\alpha+\delta)|t|-\lambda}{\alpha}.
\end{equation}

Finally, we arrive at
\begin{equation}
E_c=\frac{2\text{sgn}(\lambda+\delta \alpha)|t|-\lambda}{\alpha}.
\end{equation}

\section{Transition points by tuning  $\delta$ \label{transition point}}
Here we focus on the interval $-1<\alpha<0$ and estimate the range of energy spectrum,
while the discussion in the interval $0<\alpha<1$ is similar.
For the discussion below, the hopping amplitude $t$ is set to be $1$.
Observing the on site potential Eq.(\ref{va}), a special point is obvious: $\frac{\lambda}{\alpha}+\delta=0$.
At this point, the range of energy spectrum is $[-2,2]$ and
the eigenstates are always extended.
%because the distance between mobility edges is $|\frac{4}{\alpha}|>4$.
For convenience, we define a new parameter $\Delta=\frac{\lambda}{\alpha}+\delta$
from now on. In the following, we will discuss from two aspects.

(i) $\Delta>0$.
The energy spectrum only has cross points with the upper mobility edge line $E=-\frac{2}{\alpha}-\frac{\lambda}{\alpha}$.

When $\Delta$ is small, the approximate range of energy spectrum spectrum is $[-2+\frac{\Delta}{1-\alpha}-\frac{\lambda}{\alpha},2+\frac{\Delta}{1+\alpha}-\frac{\lambda}{\alpha}]$.
Thus, a transition point appears when the mobility edge line intersects with the energy spectrum.
It is determined by
\begin{equation}
-\frac{2}{\alpha}=2+\frac{\Delta}{1+\alpha}.
\end{equation}
So the transition point is given as
\begin{equation}
\Delta=-\frac{2(1+\alpha)^2}{\alpha}
\rightarrow \delta=-\frac{\lambda}{\alpha}-\frac{2(1+\alpha)^2}{\alpha}.
\end{equation}

When $\Delta$ is large, all the eigenstates become localized states.
In this regime, the range of energy spectrum is well approximated as $[\frac{\Delta}{1-\alpha}-\frac{\lambda}{\alpha},\frac{\Delta}{1+\alpha}-\frac{\lambda}{\alpha}]$.
And the transition point upon which all the states become localized is determined by
\begin{equation}
-\frac{2}{\alpha}=\frac{\Delta}{1-\alpha}
\end{equation}
and the transition point is
\begin{equation}
\Delta=-\frac{2(1-\alpha)}{\alpha}
\rightarrow \delta=-\frac{\lambda}{\alpha}-\frac{2(1-\alpha)}{\alpha}.
\end{equation}

(ii)  $\Delta<0$.
The energy spectrum only has cross points with lower mobility edge line $E=\frac{2}{\alpha}-\frac{\lambda}{\alpha}$.

When $|\Delta|$ is small, the approximate range of energy spectrum is $[-2+\frac{\Delta}{1+\alpha}-\frac{\lambda}{\alpha},2+\frac{\Delta}{1-\alpha}-\frac{\lambda}{\alpha}]$.
So the transition point upon which the mobility edge line meets the energy spectrum is determined by
\begin{equation}
\frac{2}{\alpha}=-2+\frac{\Delta}{1+\alpha}
\end{equation}
and the transition point is
\begin{equation}
\Delta=\frac{2(1+\alpha)^2}{\alpha}
\rightarrow \delta=-\frac{\lambda}{\alpha}+\frac{2(1+\alpha)^2}{\alpha}
\end{equation}

When $|\Delta|$ is large, all the eigenstates become localized states.
In this region, $[\frac{\Delta}{1+\alpha}-\frac{\lambda}{\alpha},\frac{\Delta}{1-\alpha}-\frac{\lambda}{\alpha}]$ is a good approximation for the range of energy spectrum.
And the transition point where all the states become localized is determined by
\begin{equation}
\frac{2}{\alpha}=\frac{\Delta}{1-\alpha}
\end{equation}
and thus the transition point given as
\begin{equation}
\Delta=\frac{2(1-\alpha)}{\alpha}
\rightarrow \delta=-\frac{\lambda}{\alpha}+\frac{2(1-\alpha)}{\alpha}.
\end{equation}

One can find that these transition points are symmetric about $\delta=-\frac{\lambda}{\alpha}$.
%However, the mobility edges in two intervals exists mobility are determined by different equations.

As $\delta$ varies, we can obtain systems which are fully localized, partially localized and fully extended.
%The transition points are also given approximately.
For intervals of $\delta$ possessing true mobility edges,
%their physical properties are also different.
it is worth noting that when $\Delta<0$, the low-energy eigenstates are localized and the high-energy eigenstates are extended, while contrarily the situation reverses when $\Delta>0$.
%Their physical properties will be different in the many body case.

%\bibliography{MyBib}


\end{document}